\documentclass[10pt, conference, compsocconf]{IEEEtran}

\usepackage{comment}  
\usepackage{graphicx}           
\usepackage{subfigure}          
\usepackage{amsmath}            
\usepackage{amssymb}            
\usepackage{url}                
\usepackage{ifpdf}              
\usepackage{pdfpages}           
\usepackage{algorithm}
\usepackage[noend]{algpseudocode}
\usepackage{multirow}

\ifpdf
  \DeclareGraphicsExtensions{.pdf,.png,.jpg}
\else
  \DeclareGraphicsExtensions{.eps,.ps}
\fi

\usepackage{float}
\floatstyle{boxed} 
\restylefloat{figure}

\begin{document}

\title{Towards a Networks-of-Networks Framework for Cyber Security}

\author{\IEEEauthorblockN{Mahantesh Halappanavar,
Sutanay Choudhury, Emilie Hogan, Peter Hui, and John R. Johnson}
\IEEEauthorblockA{Pacific Northwest National Laboratory\\
Richland, Washington 99352\\ 
Email: (First-name.Last-name)@pnnl.gov}\\  
\IEEEauthorblockN{Indrajit Ray}
\IEEEauthorblockA{Colarado State University\\
Fort Collins, Colorado\\
Email: Indrajit@cs.colostate.edu}\\
\IEEEauthorblockN{Lawrence Holder}
\IEEEauthorblockA{Washington State University\\
Pullman, Washington\\
Email: Holder@wsu.edu}
}

\maketitle

\begin{abstract}
Networks-of-networks (NoN) is a graph-theoretic model of
interdependent networks that have distinct dynamics at each network
(layer). 
By adding special edges to represent relationships between nodes in
different layers, NoN provides a unified mechanism to study
interdependent systems intertwined in a complex relationship.
While NoN based models have been proposed for cyber-physical systems,
in this {\em position paper} we build towards a three-layered NoN
model for an enterprise cyber system.
Each layer captures a different facet of a cyber system. 
We present in-depth discussion for four major graph-theoretic
applications to demonstrate how the three-layered NoN model can be
leveraged for continuous system monitoring and mission assurance.
\end{abstract}


\section{Introduction}
\label{sec.intro}
Cyber security is a tremendous challenge facing not only individual
corporations but also nations around the world. The social and
economic impact from a systematic cyber attack is potentially
devastating for an organization. Understanding and overcoming the
vulnerabilities of an enterprise cyber system is therefore critical to
continued operation of an organization. However, cyber systems are
complex in nature and hard to model in a uniform setting. A graph
$G=(V,E)$ is an ordered pair where $V$ is a set of vertices (or nodes)
and $E$ is a set of edges (or links). An edge represents a binary
relationship between two vertices. Graph is a simple but powerful tool
for modeling complex systems. Unique entities of a system form the
vertex set $V$, and binary relationships between vertices are
represented in the edge set $E$. Driven by their simplicity,
graph-theoretic modeling and analysis has been used extensively for
cyber security~\cite{R1,R2,R3,R4,R5}. Although powerful, graph-based
modeling of cyber systems has limited applicability. 
A fundamental limitation of this approach is that vertices and edges
of a graph need to be homogenous in order to make graph-theoretic
analysis meaningful. 
However, cyber systems are fundamentally heterogeneous in
nature. Graph-theoretic models are also limited in their ability to
capture the dynamics and physics of a given cyber system accurately.

To further illustrate these limitations consider the following
situation. An enterprise has access to several different sources of
information such as electronic-mail traffic within its domain, network
flow within and across its perimeter, events generated by anti-virus
software running on individual computers, and vulnerabilities of
individual computers quantified as a function of the operating system
and services that run on them. Individually, each source provides a
limited perspective on the collective actions of human and software
agents within an enterprise. By focusing on the information related to
a particular user group or a specific software service, strong
correlation or dependencies can be observed across multiple data
sources. Based on an aggregated information, addressing questions such
as: “How likely would host $h_k$ get compromised?” or “Given that host
$h_k$ has been compromised, how can we ensure continued operation of a
service $s_i$ running on $h_k$?” will become possible. It is not only
necessary to discover independent local events, but also to follow the
trail of dependencies and discover new events and vulnerabilities at a
global scale.  Therefore, we argue that it is imperative for a cyber
system model to encompass different entities (hosts, users, software
processes, and events) and their diverse interactions (user-host,
host-host, host-process, etc.).  

Further, the benefits of modeling a complex cyber system using a
graph-theoretic approach are tremendous.  However, the above
definition of a graph can only capture binary relations between
homogeneous entities. It does not provide a mechanism to distinguish
between heterogeneous entities and different types of entity
relationships. Semantic networks~\cite{R20} and conceptual graphs~\cite{R19}
support the notion of heterogeneous entities and multiple types of
edges.  However, this expressivity comes at a price.  For example,
consider a simple breadth-first search on a semantic network to detect
paths between hosts. A multi-hop path not only includes hosts but may
also include other vertex types with specific rules.  Thus, to
discover which hosts are reachable in $k$ hops we need to extend
traditional graph-theoretic tools to incorporate the notion of
heterogeneity. In summary, there is a critical need to extend
traditional graph-theoretic models to successfully model complex cyber
systems. 

Motivated by the interdependencies between different critical
infrastructure networks, the networks-of-networks (NoN) model was
developed to address some of the limitations of graph-theoretic
modeling of single networks~\cite{R9}. For instance, infrastructure networks
such as electric power grids, communication and control networks, oil
and natural gas pipelines, and transportation networks are all
interconnected and interdependent on each other. While each of these
networks can be efficiently modeled and studied individually using
graph-theoretic concepts and tools, the interdependencies are hard to
model in a uniform setting. However, the inclusion of these
dependencies fundamentally alters our understanding of a complex
network, and is therefore being labeled as the next frontier of
network science~\cite{R6}. 
Using a two-layered model, Kurant and Thiran~\cite{R7}
made a distinction between the physical and the logical aspects of a
network. A mapping of the logical network onto the physical network
was then added. This approach enabled a better modeling of the load on
a (railroad) network and the vulnerabilities caused due to failures in
different layers. Further, the notion of NoN naturally emerges in the
context of cyber-physical systems. For example, the interactions
between a supervisory control and data acquisition (SCADA) network and
the underlying power transmission network that SCADA controls. The
connectivity and dynamics of these two networks are fundamentally
different, but the two networks are interdependent on each other. By
modeling this system as a NoN critical properties and combined
vulnerabilities can be expressed mathematically~\cite{R8,R9}. 

{\bf Contributions}: In this {\em position} paper, we propose a novel
three-layered NoN model for an {\em enterprise} cyber system (Section
!\ref{sec.basic_model}). 
By explicitly modeling the different facets of a cyber system,
our goal is to bring together different concepts from traditional
graph theory and network science to bear on the problems in cyber
security (Section~\ref{graph_analysis}). 
We aim to develop an efficient tool for modeling complex cyber systems
and provide graph-theoretic metrics to express different
functionalities and vulnerabilities of such a system. 
While NoN models have been proposed for several cyber-physical
systems, our approach is to apply this model exclusively for cyber
systems, and is thus the novelty of our approach.

\section{A NoN-Based Model for Cyber Systems}
\label{sec.basic_model}
\subsection{Preliminaries}
\label{subsec:defs}
Mathematically, we represent a network of $l$ networks as a graph
$G=(V,E\cup E^{'})$ where the vertex-set $V=V_1\cup V_2 \ldots \cup
V_l$, and the edge-sets $E=E_1\cup E_2\ldots \cup E_l$ and
$E^{'}=E_{1,2}\cup E_{1,3} \ldots \cup E_{i,j}$, such that $V_i \cap
V_j=\emptyset$, and $E_i \cap E_j =\emptyset$ for $i,j=1\ldots l$. The
graph can have weights associated with vertices and/or edges $w:E\to
\mathbf{R}$, and can be directed or undirected depending on the
network that is being modeled. Pair-wise interconnections between two
networks can be represented as a bipartite graph $G^{'}=(V_s\cup V_t,
E_{s,t})$, such that $V_s \cap V_t=\emptyset$ and an edge always
connects a vertex in $V_s$ to a vertex in $V_t$. Further, we define an
edge $e \in E$ between two vertices $(i,j)$ with its end-points as
$e_{i,j}$ where appropriate. 
We define a walk in a graph as a finite sequence of edges (or
vertices) $W=\{e_{1,2}, e_{2,3}, e_{3,4} \ldots e_{k-1,k}\}$ 
such that any two consecutive edges in $W$ are adjacent to each other
in the graph. A walk that does not traverse any edge twice and visits
each vertex only once is called a path. We now briefly present a
three-layered NoN model to represent an enterprise cyber system.

\subsection{Three Layered Model}
For the purposes of representing an enterprise cyber system, we
propose a three-layered NoN model, where the layers are: 
\begin{itemize}
\item 	Layer-1: Physical (Hardware) layer,
\item 	Layer-2: Logical (Software; Functional) layer, and
\item 	Layer-3: Social (User; Computer) layer.
\end{itemize}

This simple model, illustrated in Figure 1, can be further extended to
include other aspects of an enterprise as needed. For example, modern
telecommunication systems critically depend on the underlying
communication networks for operation – this can be modeled as a new
layer. Another example would be the network of power cables supporting
the communication network, which forms a new layer. We now provide
details on how to model each layer and to build their
interdependences. 

\begin{figure}
\label{fig:NoN}
\includegraphics[width=3.5in]{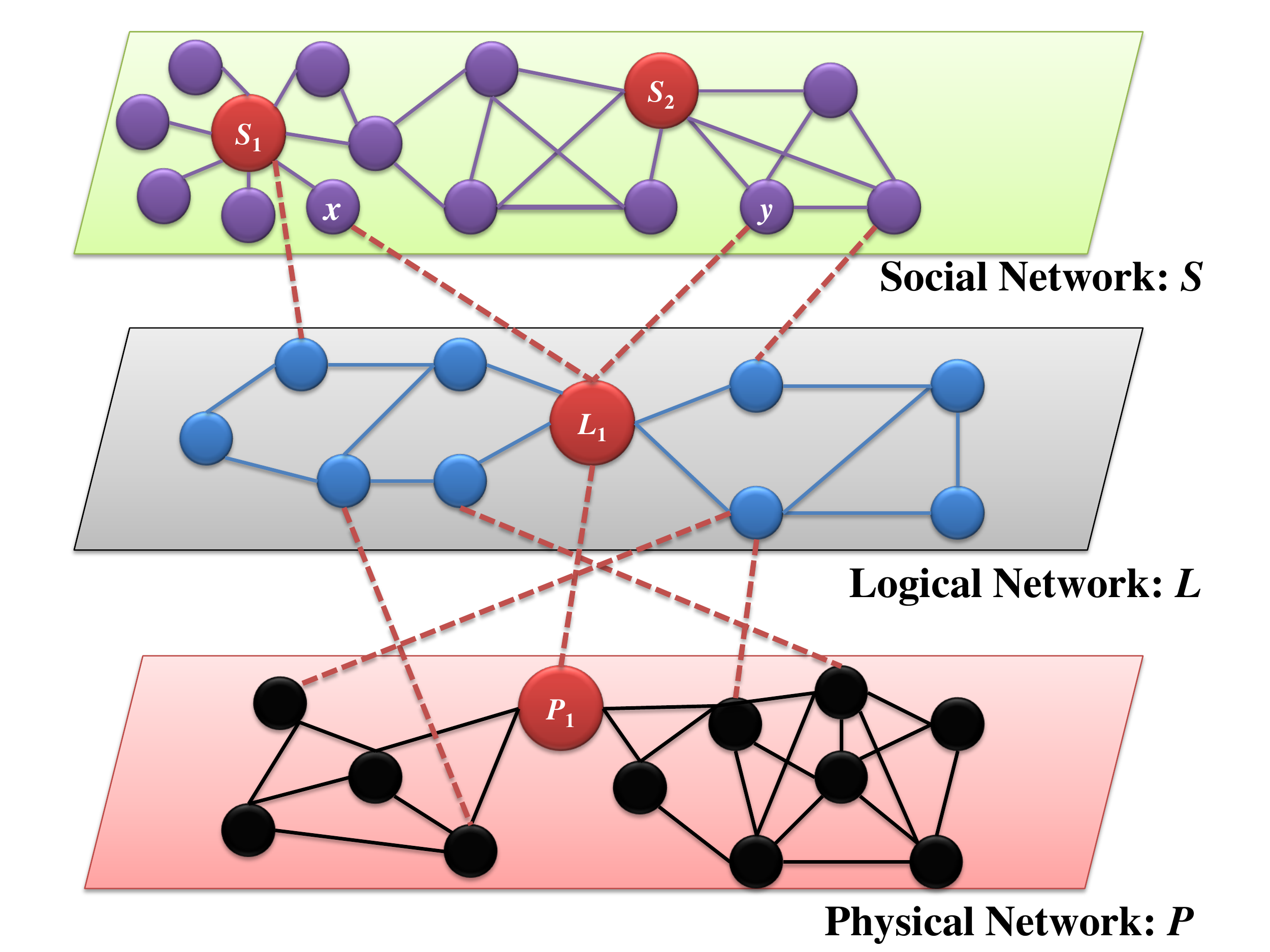}
\caption{\label{case30}
An illustration of a three-layered (social--logical--physical)
network-of-networks model.} 
\end{figure}

The physical layer (Layer-1) is built from the network topology that
is readily available to network engineers and administrators. This
network consists of computers that are interconnected via switches and
routers to the enterprise backbone. From a graph-theoretic perspective
the connectivity of this network is fundamentally different from other
layers. For example, let different workstations (computers) be
represented as vertices, and two computers that can connected by a
certain physical link be represented by edges. In this representation,
all the computers connected to a switch would form a clique or a
strongly connected component (a path exists from each node to every
other node in the component). If we treat virtual machines as
independent servers (vertices), then all the virtual machines residing
on a physical server would also form a clique. 
  
The logical layer (Layer-2) can be built in several different
ways. One simple approach would be to model the different software
applications (services) running in an enterprise system as a
graph. Here the nodes would correspond to different services and edges
would represent pair-wise relationships between them. Other approaches
such as attack graphs [3] that describe vulnerability dependencies and
system exploits would also form a network for the logical
layer. Similar to the approach taken by Karagiannis {\em et al.}~\cite{R5},
information collected from analyzing NetFlow~\cite{R16} data can lead to the
functional relationships between different services. When a logical
network is built to represent different services, the relationship
with the physical layer can be straightforward – services run on
specific hardware. On the other hand, if relationships are not obvious
simplifying assumptions can be made. From a graph-theoretic
perspective the connectivity structure of networks at this layer is
different from those at the physical layer. For example, attack graphs
are typically a collection of directed paths from an attack source to
an attack destination (target computer). 

The social layer (Layer-3) can be built in several different ways. Our
goal at this abstraction layer is to capture the social aspects of a
network:  How do we find users with similar activity-profile? How do
different users within the enterprise interact? The interactions among
different users can be collated from several sources such as
electronic mail exchanges, instant messenger communications and
possibly through participation in social networks. Similarly, social
interaction of computers via their Internet Protocol (IP) addresses
can be collated from network monitoring tools such as NetFlow. Graphs
describing behavioral similarity can be built by clustering the user
activities. An edge is added between two users/hosts if both of them
participate in the same cluster by satisfying a threshold on
membership. From a graph-theoretic perspective such networks have been
studied extensively~\cite{R11}. 

In general, social interaction networks tend to have the following
structural properties: $(i)$ small-world nature, where any two nodes
in the graph are connected via a short geodesic path; $(ii)$ large
clustering coefficient, where the neighbors of a node are also
interconnected; and $(iii)$ scale-free nature, where a large number of
nodes have a small degree (number of neighbors) and a small number of
nodes have large degrees. As a consequence, social networks tend to be
vulnerable to targeted attacks but resilient to random failures. 

\subsection{Interactions between layers}
Following the work of Buldyrev et al. we use special edges (dependency
links) to connect different networks between layers [8, 9]. Dependency
edges are directed edges. Further, we make a distinction between
dependent and independent edges. Independent edges represent
internetwork relationships where the node in one network does not
depend on the node in another network for operation. Independent edges
are modeled as undirected (bidirectional) edges. This distinction will
become clear for a given context and the graph algorithms that are
being used, which will be discussed next.

\section{Graph-Theoretic Analysis for Mission Assurance}
\label{graph_analysis}
Our goal is to study and quantify different aspects of an enterprise
cyber system. Globally at an enterprise level, we are interested in
understanding and quantifying the collective robustness (or
vulnerability) of a cyber system due to targeted or random failures at
different layers (networks). And, locally at a subsystem level, we are
interested in developing metrics that can be computed from information
collected locally but provide enough information to guide continuous
monitoring and maintenance. The rest of this section is dedicated to
answering the above questions using the NoN model.  We present a
number of graph-theoretic approaches that measure and describe
different facets of a cyber system.  

\subsection{Critical nodes and their relationships}
Identifying critical components (represented as nodes and/or edges of
a graph) in a network is an important activity. Many graph-theoretic
approaches can be used to identify such components. For example,
betweenness centrality~\cite{R21} measures the importance of a node (or an
edge) as the ratio of the number of paths (for example, shortest paths
as described in Section~\ref{subsec:defs}) that flow through a
particular node to the number of all possible paths in the network. 
If the ratio is high for a given node then it implies that this node
is critical to the connectivity of the graph. 
While betweenness centrality is one of the
popular measures, there are several other variants of centrality that
quantify different aspects of connectivity in a graph~\cite{R12}. Metrics
based on network flow, nodes along a maximum-flow minimum-cut in a
graph, have also been used in the context of cyber security to
identify critical nodes~\cite{R2}. In the context of networks-of-networks,
computation of these metrics needs to be adapted carefully. As an
illustration, consider the bottom two layers (logical-physical) of the
three-layer NoN model illustrated in Figure 1. In the logical network,
$L_1$ is an important node (since it controls the flow from the left
part of the network to the right). Similarly, in the physical network,
$P_1$ is an important node. If node $L_1$ is dependent on any node in
$P$, then the failure of node in $P$ will be detrimental to the
connectivity of the logical network. Therefore, identifying critical
nodes and their interrelationships is an important activity towards
making a cyber system robust.  

Quantifying the relationships between critical nodes is an important
next step. Let us assume that we have identified critical nodes in
different layers based on their degree (number of neighbors)
distributions. In the context of a single network, Newman defined the
concept of assortativity as the association of nodes that are similar
in some fashion, degree for example~\cite{R13}. Similarly, a network is
called a disassortative when dissimilar nodes get associated with each
other. Assortativity is measured as Pearson correlation coefficient of
the degrees of nodes at the two ends of edges. The metric is positive
for assortative networks, negative for disassortative networks, and
zero for networks with no specific associations. The notion of
assortativity can be extended to networks-of-networks and provides a
critical metric to gain insight to the interdependencies. Parshani
{\em et al.} define this as the joint probability of a dependency edge
connecting a node in the first network with degree j to a node in the
second network with degree k, and call it inter degree-degree
correlation (IDDC). Using simulations and empirical data modeling a
two-lawyer port-airport system, they show that coupled systems with
higher assortativity between layers are more robust to random failures
~\cite{R10}. While vertex degrees are important, we need the ability to use
other metrics to define assortativity in cyber systems. The notion of
assortativity needs to be further modified for NoNs where we might get
a zero assortativity when non-critical nodes in one network are
connected to critical nodes in another. Another closely related metric
of interest for us is the notion of clustering coefficient which can
be extended to networks-of-networks. Local clustering coefficient of a
vertex in a graph is defined as the ratio of the number of actual
edges to the total number of potential edges between all neighbors of
this vertex. Parshani {\em et al.} extend this to NoN as the ratio of
the number of neighbors of a node in the first network that are also
dependent on the corresponding neighbors of its dependent node in the
second network. They call this metric as the inter-cluster coefficient
(ICC)~\cite{R10}. Similar to IDDC, this metric provides insight on the
collective robustness (vulnerability) of NoN. 

\subsection{Reachbility Analysis}
\label{subsec:reach}
Reachability analysis can be a useful tool in determining the
vulnerability of a network. If it is important that a chosen vertex
$v$ in $L$ (in Figure 1) be reachable quickly from another vertex $w$
also in $L$, then we can count the number of shortest paths (as
defined in Section~\ref{subsec:defs}) between $v$ and $w$ in $L$. However, it
might be possible that the shortest path between two vertices in a NoN
model may go through other layers (consider vertices $x$ and $y$ in
layer $S$ from Figure 1). Disturbances in one layer would result in
the loss of efficient connectivity in the other (the dependent
layer). On the other hand, the fact that there are paths between $v$
and $w$ that go through $L$, and paths that go through $P$ can be
positive in the sense that the network as a whole becomes more
resilient. Thus, by suitably expressing reachability across multiple
layers, we can measure robustness as a function of reachability. Using
a single-network approach we used reachability as a metric to express
vulnerability of a cyber system~\cite{R23}, and plan to extend this approach
to multi-layer networks in our future work. 

\subsection{Cascading Failures}
A topic that originated in the context of electric power grids but has
found applications in a wide range of networks is the concept of
cascading failures, popularly known as blackouts in power grids. The
dynamics of network failures can be modeled in several ways. Cascading
failures have been studied using different models under different
names. We will extend a few models that are relevant to NoN based
models of an enterprise cyber system. A commonly studied topic under
this theme is of graph perturbation where changes are made to a graph
and the impact of these changes is studied~\cite{R14}. Different metrics can
be used to measure the impact of node (or edge) failures: connectivity
of the graph expressed using the number of disconnected components, or
the size of the largest connected component, or the fraction of nodes
remaining in the largest connected component; flow in the network; or
the overall modifications to the size of the network expressed via the
number of nodes and/or edges.  

Another analysis that will be useful for addressing aspects of
cascading failures is optimal security hardening~\cite{R24}. In order to
protect a system from cyber breaches, all known weaknesses in a system
needs to be hardened. However, there is an associated cost to this and
a fixed budget dictates which security controls can be implemented and
which weaknesses need to be left unhardened. Thus, first and foremost,
an optimal strategy needs to be determined that minimizes residual
damage (damages due to having weaknesses in a system) to the NoN that
is not perfect. In order to perform such analysis, we can model the
NoN as a dynamic graph and focus on the degree distribution of the
nodes in the NoN as it changes over time. Of most importance will be
nodes that have high centrality – the higher the centrality the more
valuable a node is. Starting with nodes with high centrality, further
node reachability questions can be asked that will allow us to
identify critical nodes that potentially contribute to cascading
failures and design efficient defense strategies to overcome
them. This information can then be used in a multi-objective
optimization problem~\cite{R25} that minimizes the total security control
cost as well as minimizes the residual damage. We can constrain this
optimization problem by allowing a maximum degree of perturbation in
the NoN model. 

\subsection{Subgraph Pattern Mining}
Evidence of a cyber attack can be extracted from specific signatures
expressed as subgraphs or graphlets. Therefore, the ability to detect
and count a large number of signatures from cyber data is
desirable. Detection of emerging patterns can be used to counter
active attacks. Kargiannis {\em et al.} use this approach for
classification at the application level~\cite{R5,R22}. However, we can use
the graphlet degree distribution (GDD) proposed by Pr\v{z}ulj~\cite{R15} as a
general metric to study the existence of patterns not only in the
static data but also to express similarity between two networks.  

While the identification of specific signature subgraphs and graphlets
can provide insight into the behavior represented by the NoN, mining
for more general subgraph patterns and anomalies has shown promise for
cyber-security applications~\cite{R17,R18}. Incorporating multiple views of
the domain (physical, logical, social) into one graph potentially
allows for the discovery of rich patterns involving structure from
multiple views; however, such approaches would not scale to the size
and rate of change over the entire NoN. Mining within layers will be
more efficient, because an individual layer is smaller and more
homogeneous. But more importantly, subgraph patterns and anomalies in
one layer serve to focus attention in other layers. For example, a
frequent subgraph pattern in the social layer can be expanded to
include the induced subgraph from the logical layer to improve
understanding of the services supporting the normative social
behavior. Likewise, an anomaly in the social layer gives us direct
links to the related structures in the logical and physical layers,
thus providing an explanation of lower-level anomalies. And the
reverse forensics can link patterns and anomalies in the physical or
logical layers to the responsible entities in the social layer. The
NoN also improves our ability to mine for patterns in the presence of
dynamics, because change in the structure of one layer can be
processed independently from the change in another. This allows us to
improve graph mining efficiency while still retaining the connections
between layers in order to perform a more focused analysis of the
dynamics in other layers.

\section{Summary and Future Work}
\label{sec:summary}
We proposed a novel networks-of-networks (NoN) approach for modeling
an enterprise cyber system. We claim that such an approach enables
efficient modeling of interactions between heterogeneous entities and
yet allows us to leverage on the traditional graph-theoretic tools. We
presented a novel three-layered model to demonstrate how the NoN
approach can be used for continuous system monitoring and mission
assurance.  Specifically, we discussed in depth the following
graph-theoretic methods for cyber security: $(i)$ detection of
critical nodes and their dependencies; $(ii)$ reachability analysis;
$(iii)$ models for cascading failures; and $(iv)$ subgraph pattern
mining. We described how each of these applications can be built on
top of our proposed NoN model, thus extending the state-of-the-art.  

Building a framework to extract each of the unique networks from
Netflow, electronic mail and event logs is the next logical step for
our work. Discovering dependencies across these networks and
developing theories to translate these dependencies into answering
questions on risk management and mission assurance remain candidates
for our future work. NoN is an emerging novel concept and we look
forward to exciting research in the near future.

\section*{Acknowledgment}
This work is funded by the Asymmetric Resilient Cyber (ARC) Initiative
at the U.S. Department of Energy’s Pacific Northwest National
Laboratory (PNNL). 
PNNL is operated by Battelle Memorial Institute under Contract
DE-ACO6-76RL01830. 
We thank John Feo, Pradeep Ramuhalli and George Muller for stimulating
discussions on this topic. 


\bibliographystyle{IEEEtran}
\bibliography{references}

\end{document}